\documentclass{article}
\usepackage{ais25}
\usepackage[T1]{fontenc}
\usepackage{graphicx}
\usepackage{url}
\usepackage{color}
\usepackage[hidelinks]{hyperref}
\usepackage{cleveref}
\usepackage{tcolorbox}

\crefformat{section}{\S#2#1#3}
\crefformat{subsection}{\S#2#1#3}

\title{The Quest for Reliable Metrics of Responsible AI}
\author{\name{Theresia Veronika Rampisela}  \addr{University of Copenhagen} \email{thra@di.ku.dk} \\
\name{Maria Maistro} \addr{University of Copenhagen}  \email{mm@di.ku.dk} \\
\name{Tuukka Ruotsalo} \addr{University of Copenhagen and LUT University}  \email{tr@di.ku.dk} \\
\name{Christina Lioma} \addr{University of Copenhagen}  \email{c.lioma@di.ku.dk}}

\begin{document}

\maketitle

\begin{abstract}
The development of Artificial Intelligence (AI), including AI in Science (AIS), should be done following the principles of responsible AI. Progress in responsible AI is often quantified through evaluation metrics, yet there has been less work on assessing the robustness and reliability of the metrics themselves.
We reflect on prior work that examines the robustness of fairness metrics for recommender systems as a type of AI application and summarise their key takeaways into a set of non-exhaustive guidelines for developing reliable metrics of responsible AI. 
Our guidelines apply to a broad spectrum of AI applications, including AIS. 
\end{abstract}


\section{Introduction}

Recent legislation,  
such as the EU Artificial Intelligence (AI) Act and the EU Digital Services Act, has increasingly emphasised the responsible development of AI applications \citep{EU_DSA_2022,EUAIAct2024} to protect their users and minimise harms. 
One such AI application is Recommender Systems (RSs). RSs are highly useful in  finding relevant items that match user needs, preferences, or interests \citep{AggarwalRecommenderSystems}. These systems are widely used for both high- and low-stakes activities in professional and personal settings. 
In professional settings, among other use cases, RSs can be used to suggest scientific papers and citations for researchers \citep{Kreutz2022Scientific,Farber2020Citation}, or job positions for jobseekers \citep{Siting2012JobSurvey}.

Prior work has investigated aspects of responsible AI in RSs, such as fairness and bias \citep{Wang2023ASystems,Klimashevskaia2024SurveyPopularityBias}. Biases may exist in the data, the algorithm, or the evaluation pipeline of an RS; all these biases can contribute to unfairness. 
While there is no universal definition, fairness in RSs is commonly understood as providing equal treatment or equitable outcomes to the RS stakeholders \citep{Wang2023ASystems}, ensuring that no individuals or groups are systematically discriminated. 

Unfair RSs can have severe real-life consequences.
In job recommendations, for example, unfair RSs may contribute to the exacerbation of gender pay gaps. This could happen if historically marginalised groups (e.g., women) are only shown recommendations for lower-paying jobs, while highly-paid positions are only recommended to the historically dominant group. In the AIS context, an unfair paper/citation RS could perform extremely well in recommending relevant work for one discipline (e.g., computer science) but perform poorly for other disciplines (e.g., Nordic studies) due to training data imbalance between various fields of study, hindering the development of sciences. Similarly, if an unfair RS overpromotes articles by researchers from economically developed countries \citep{trec-fair-ranking-2020}, it may simultaneously provide less exposure to researchers from other countries. This imbalance could lead to a less inclusive understanding of science \citep{Mihalcea2025WEIRD}, especially in social sciences and humanities, where cultural context in particular matters. Hence, the development of fair RSs is not only scientifically relevant but also societally important.

Many evaluation approaches and metrics have been proposed to measure RS fairness. Yet, few studies have investigated the reliability and robustness of these approaches. This leads to the metric score being potentially misleading or unstable, and thus unreliable. Having reliable metrics is critical as they primarily guide the responsible development of AI(S) applications, especially in the early stages, as they provide a scalable way of evaluating system performance without significant cost (e.g., opposed to user studies).
In this paper, we condense insights from our previous work on RS fairness evaluation \citep{Rampisela2024EvaluationStudy,Rampisela2024CanRelevance,Rampisela2025RelevanceGuidelines,Rampisela2025StairwayFairness} into a set of guidelines for developing reliable metrics of responsible AI, based on the limitations that we uncovered in existing approaches. As these guidelines are general, they would also be relevant for evaluating AI in science (AIS). 

\section{Fairness Evaluation}

In this part, we explain how fairness, an aspect of responsible AI, is evaluated in RSs  (\Cref{ss:eval_fair_rs}). We summarise our findings and contributions from previous work on RS fairness evaluation (\Cref{ss:findings}).

\subsection{Evaluation of Fairness in Recommender Systems}
\label{ss:eval_fair_rs}

At a high level, Recommender System (RS) fairness evaluation is often categorised in terms of the \textit{subject} (user/item fairness) and the evaluation \textit{granularity} (group/individual fairness) \citep{Wang2023ASystems}. \textit{User fairness} concerns the disparity in the recommendation effectiveness, where an RS that performs equally well for all users would be deemed fairer than an RS that can perform very well for some users but fails to provide relevant recommendations for others. In contrast, \textit{item fairness} is usually defined in terms of how much exposure is received by the item, in comparison to other items in the RS. 
In terms of granularity, \textit{group fairness} concerns the difference in utility (recommendation effectiveness or item exposure) between subject groups, where the groups are typically formed based on user or item attributes, e.g., socio-demographic attributes. Meanwhile, \textit{individual fairness} is frequently operationalised as the variation of utility across all users or items.

RS practitioners often rely on computing metrics as a proxy for fairness. Over 30 metrics have been used to measure fairness for different subjects and at different granularity levels \citep{Wang2023ASystems}. 
Some metrics are based on well-known inequality measures, such as standard deviation, entropy, and Gini Index. Others are newly developed based on existing fairness notions, e.g., Rawlsian fairness and envy-freeness. With the sheer amount of metrics, it is often difficult to understand what and how they actually quantify fairness. Even worse, some metrics were proposed without further analysis of their limitations. Consequently, the measure limitations are largely unknown.

\subsection[]{Our Contributions to RS Fairness Evaluation}
\label{ss:findings}

We analysed several RS fairness metrics and uncovered limitations within the measures. To resolve the limitations, we proposed novel fairness evaluation approaches. Finally, we provided practical guidelines for evaluating RS fairness.

\paragraph{Analysis on metric limitations.}
Our investigation on metric limitations was done both theoretically and empirically.
In our theoretical studies, we show that some fairness metrics are mathematically flawed \citep{Rampisela2024EvaluationStudy,Rampisela2025RelevanceGuidelines}. 
Firstly, some metrics would either crash upon computation due to invalid mathematical operations (e.g., division by 0), resulting in the metric outputting no scores that can be used for fairness assessment, rendering it unusable. 
Secondly, a large number of the metrics have an unknown score range, or if it is known, the max/min score is not reachable. Suppose that in theory, a metric score ranges between 0 and 1, where 0 is the fairest score and 1 is the unfairest score. In some cases, there is no input to the metric that would output  0 or 1. Instead, for example, its score could only range between 0.3 and 0.6 (i.e., the minimum and maximum reachable scores, respectively). This causes difficulty in interpreting the metric score, as one would interpret a score of 0.5 as somehow fair if the range is $[0, 1]$, but unfair if the range is $[0.3, 0.6]$. 
Thirdly, in some cases, it is not even known what kind of input to the metric would result in the max/min scores, resulting in a limited understanding of what the (un)fairest possible case is, according to the metric.

Our empirical studies showed that some metrics tend to score very low (close to 0) independently of the fairness level. The compressed score range and limited metric sensitivity give the illusion of an extremely fair input, even when it is not fair, misleading the understanding of how (un)fair a system is. Additionally, we found that some metrics may be redundant as they yield similar conclusions to other existing metrics. As such, computing one of the similar metrics may suffice. On the other hand, we showed that metrics for one granularity level cannot be used as a proxy to the other, i.e., group fairness metrics cannot estimate individual fairness \citep{Rampisela2025StairwayFairness}. 
Therefore, it is necessary to consider different granularities to have a more comprehensive view of fairness.

By investigating the metrics' limitations, we hope to raise awareness that a metric could suffer from several issues that could affect its usability and interpretability. Thus, one should not immediately trust a metric's score at face value without truly understanding how the metric operates.

\paragraph{New evaluation approaches.} 
We contributed two types of new evaluation approaches: metric corrections and new metrics. 
Firstly, we correct existing fairness metrics by redefining their formulation to avoid computation crash due to invalid mathematical operations. We also applied min-max normalisation, such that the score 0 is correctly mapped to the fairest possible case and 1 to the unfairest possible case \citep{Rampisela2024EvaluationStudy, Rampisela2025RelevanceGuidelines}. The corrected metrics can then be interpreted more easily than the original metrics. Secondly, we proposed a new metric to jointly evaluate recommender system effectiveness and fairness, as existing metrics cannot quantify both aspects simultaneously \citep{Rampisela2025JointFrontier}. 
We have also released the source codes to compute both new evaluation approaches publicly, so that they can be used more easily.\footnote{For the corrected metrics, please refer to  \href{https://github.com/theresiavr/individual-item-fairness-measures-recsys}{github.com/theresiavr/individual-item-fairness-measures-recsys} and \href{https://github.com/theresiavr/relevance-aware-item-fairness-measures-recsys}{github.com/theresiavr/relevance-aware-item-fairness-measures-recsys}. For the effectiveness-fairness joint evaluation approach, please refer to \href{https://github.com/theresiavr/DPFR-recsys-evaluation}{github.com/theresiavr/DPFR-recsys-evaluation}.}

\paragraph{Practical guidelines.}

Considering that there are many RS fairness metrics with their own limitations, we summarise practical guidelines for selecting the metrics from the findings in our previous work. Firstly, we recommend using our corrected metrics to evaluate how close a recommendation is to the fairest recommendation scenario, as the original metrics cannot be used to do so. Secondly, the metric scores should be interpreted carefully, as some measures tend to score very close to the fairest case and may overestimate fairness. Thirdly, the use of redundant (highly similar) measures should be avoided. Fourthly, one should evaluate for both group and individual fairness to ensure that no groups or individuals are disadvantaged.

\section{Guidelines for Formulating Reliable Metrics}
\label{ss:guidelines}

With the rise of AI usage in science, it is inevitable that new metrics may be formulated to evaluate their performance, including how well they align with responsible AI aspects. Reflecting and generalising on insights from our work on RS fairness evaluation, we provide a set of questions to guide the development of reliable metrics for responsible AI: 
\begin{enumerate}
    \item Are there input cases that should be excluded, such that the metric does not have invalid mathematical operations?
    \item What is the metric range and how should it be interpreted?
    \item What kind of input results in the minimum and maximum metric score?
    \item How sensitive is the metric to changes in the input? 
    \item Does the metric yield a similar conclusion to an existing one?
\end{enumerate}

\section{Conclusions}

Our guidelines are not meant to be exhaustive. Rather, these are the bare minimum to ensure reliable metrics for AI technologies. We intend to raise awareness, so that effort should be put not only into developing new technology, but also into improving its evaluation. Quantification is an important part of regulation and policy-making, yet evaluation metrics are frequently absent from the current AI policy discussion.\footnote{\url{https://oecd.ai/en/catalogue/overview}} As such, future work could potentially involve collaboration with AI users, technical actors, social scientists, experts in ethics, policy makers, and governmental agencies to discuss the appropriate responsible AI evaluation metric(s) to incorporate in AI policies and regulations, especially in high-stakes contexts such as AIS. The collaboration could potentially study which responsible AI aspects are critical for AIS and perform an audit of existing evaluation metrics to ensure their reliability in measuring the intended aspects. We expect the collaboration to provide more concrete, measurable guidelines for the responsible development of AI applications.

\bibliography{references}

\end{document}